\documentclass[12pt]{article}

\newenvironment{numberedlist}
{\begin{list}{\makebox[20pt]{\hss(\arabic{itemno})\enspace}}
             {\usecounter{itemno}\labelwidth 20pt}}{\end{list}}

\newcounter{itemno}

\newcounter{itemno1}

\newcounter{itemno2}
\newcounter{lemma}
\newcounter{exno}

\newcounter{defno}

\newenvironment{defn}{\refstepcounter{defno}\medskip \noindent {\bf
Definition \thedefno.\ }}{\medskip}

\newcommand{\sep}{\;\vert\;}

\newcommand{\oprove}{\vdash\kern-.6em\lower.7ex\hbox{$\scriptstyle O$}\,}

\newcommand{\Pscr}{{\cal P}}

\newcommand{\pderivation}{{\cal P}\kern -.1em\hbox{\rm -derivation}}
\newcommand{\pderivationl}{{\cal P}\kern -.1em\hbox{\em -derivation}}
\newcommand{\pderivable}{{\cal P}\kern -.1em\hbox{\rm -derivable}}
\newcommand{\pderivablel}{{\cal P}\kern -.1em\hbox{\em -derivable}}
\newcommand{\pderivations}{{\cal P}\kern -.1em\hbox{\rm -derivations}}
\newcommand{\pderivability}{{\cal P}\kern -.1em\hbox{\rm -derivability}}

\newcommand{\all}{\forall}
\newcommand{\some}{\exists}

\newcommand{\ie}{{\em i.e.}}

\newsavebox{\lpartfig}
\newsavebox{\rpartfig}

\newenvironment{exmple}{
 \begingroup \begin{tabbing} \hspace{2em}\= \hspace{3em}\= \hspace{3em}\=
\hspace{3em}\= \hspace{3em}\= \hspace{3em}\= \kill}{
 \end{tabbing}\endgroup}

\newcommand{\lb}{\langle}
\newcommand{\rb}{\rangle}
\newcommand{\pr}{prov}

\newcommand{\prove}{exec} 
  
\newcommand{\add}{\oplus} 
\newcommand{\adc}{\&} 

\newtheorem{theorem}[lemma]{Theorem}

     {\\* \hspace*{\fill} \end{trivlist}}

\newcommand{\muprolog}{{Prolog$^{\add,\adc}$}}
\renewcommand{\pr}{pv}
\renewcommand{\prove}{ex} 

\begin{document}

\begin{center}
{\Large {\bf Towards Interactive Logic Programming}}
\\[20pt] 
{\bf Keehang Kwon and Mi-Young Park}\\
Dept. of Computer  Engineering, DongA University \\
Busan 604-714, Korea\\
 \{ khkwon,openmp \}@dau.ac.kr\\
\end{center}

\noindent {\bf Abstract}: 
Adding interaction to  logic programming is an essential task. 
 Expressive logics such as linear logic provide a theoretical basis for such a
 mechanism.
 Unfortunately, none of the existing (linear) logic languages can model interactions with the user because
they uses provability as the sole basis for computation.
In the operational semantics based on  provability, 
executing the additive-conjunctive goal $G_0 \adc G_1$ from a program $\Pscr$ 
simply terminates with a success if both $G_0$ and $G_1$ are solvable from $\Pscr$.
This is an unsatisfactory situation, as  a central action of $\adc$ -- 
the action of  choosing either $G_0$ or $G_1$ by the
user -- is missing in this  semantics.

   We propose to modify the  operational
semantics above to allow for more active participation from the user. We illustrate our idea
via \muprolog, an extension of Prolog with additive goals.

{\bf keywords:} interaction, logic programming, linear logic, computability logic.



\section{Introduction}\label{sec:intro}

Adding interaction to logic and logic programming is a challenging and an essential task. An interactive logic 
program must be able to allow
the user to select one among many alternatives. Expressive logics such as linear logic provide a theoretical basis for such a
 mechanism.

Unfortunately, none of the existing (linear) logic languages can model decision steps from the user in the course of execution.
 This deficiency is an outcome of
using  provability as the sole basis for computation.
In the operational semantics based on  provability such as  uniform provability 
\cite{HM94,Mil89jlp,MNPS91}, 
solving the additive-conjunctive goal $G_0 \adc G_1$ from a program $\Pscr$
 simply {\it terminates} with a success if both $G_0$ and $G_1$ are solvable from $\Pscr$.
This  semantics, $\pr$ ($n$ stands for noninteractive), is shown below:

\[ \pr(\Pscr, G_0\adc G_1)\ if\ \pr(\Pscr, G_0)\ and\ \pr(\Pscr, G_1) \] 

\noindent
This is an unsatisfactory situation, as  an action of choosing either $G_0$ 
or $G_1$ by the
user --  the declarative reading  of $\adc$ -- is missing in this operational semantics.

 Our approach in this paper involves a modification of the  operational
semantics to allow for more active participation from the user.
Executing the additive-conjunctive goal $G_0 \adc G_1$ from a program $\Pscr$ now  has the
following operational semantics:

\[ \prove(\Pscr, G_0\adc G_1)\ if\ read(k)\ and\ \prove(\Pscr, G_i)\ and\   \pr(\Pscr, G_j) \] 

\noindent where  $i\ (=\ 0\ {\rm or}\ 1))$ is the value stored in $k$ and $j$ is $(j+1)\ mod\ 2$.
In the above definition, the system  requests the user to choose $i$ and then proceeds
with solving both the chosen goal, $G_i$,  and the unchosen goal, $G_j$.
Both executions must succeed for the current goal to succeed.
It is worth noting that
 solving the unchosen goal, $G_j$, must proceed using $\pr$ rather than $\prove$
to ensure that
there will be no further interactions  with the user.
It can be easily seen that our new semantics has the advantage over the old semantics:
the former respects the declarative reading of $\adc$
without losing completeness or efficiency.

As an   
illustration of this approach, let us consider a fast-food restaurant where you can have the
hamburger set or the fishburger set. For a hamburger set, you can have a hamburger, a coke and
a side-dish vegetable (onion or corn but they make the choice).
 For a fishburger set, you can 
have a fishburger, a coke and a side-dish vegetable (onion or corn but they make the choice). 
This is provided by the following definition:

\begin{exmple}
$! hburger.$\\
$! fburger.$\\
$! coke.$\\
$! onion.$\\
$! (hset$ ${\rm :-}$ \> \hspace{2em}      $hburger \otimes coke\ \otimes (onion \add\ corn))$\\ 
$! (fset$ ${\rm :-}$ \> \hspace{2em}      $fburger \otimes coke\ \otimes (onion \add\ corn))$\\
\end{exmple}
\noindent Here, ${\rm :-}$ represents reverse implication.
The  definition above consists of  reusable resources, denoted by $!$.
 As a particular example, consider a goal task $hset \adc fset$. This goal simply terminates
with a success in the context of \cite{HM94} as both goals are solvable.  However, in our context,
  execution proceeds as follows: the system 
 requests the user to select a particular burger set. 
After the set -- 
 say, $hset$ -- is selected, the system tries to solve the first conjunct using the new semantics, whereas it 
tries to solve the second conjunct using the old semantics.
 Now the execution terminates with a success, as both conjuncts are solvable.

     As seen from the example above, additive-conjunctive  goals can be used to model 
interactive decision tasks. We also adopt additive-disjunctive goals which are 
   of the form $G_0 \add G_1$ 
 where $G_0, G_1$ are goals.
Executing this goal has the following intended semantics: select the true disjunct
$G_i$ and execute $G_i$ where $i (= 0\ {\rm or}\ 1)$ is chosen by the system.

   To present our idea as simple as possible, this paper focuses on \muprolog,
 which is a variant of a  subset of Lolli\cite{HM94}. 
The former can be obtained
from the latter by (a) disallowing linear context and $\adc$ in the clauses, and (b) allowing only
          $\otimes, \oplus, \&$ operators   in goal formulas. \muprolog\ can also be seen
as an extension of Prolog with $\oplus, \&$ operators   in goal formulas, as
$\otimes$ in \muprolog\ corresponds to $\land$ of Prolog.

In this paper we present the syntax and semantics of this extended language, 
show some examples of its use. 
The remainder of this paper is structured as follows. We describe  \muprolog\
based on a first-order  clauses  in
the next section and Section 3. In Section \ref{sec:modules}, we
present some examples of  \muprolog. 
Section~\ref{sec:conc} concludes the paper. In Section 5, we
present an altenative execution model.

\section{The \muprolog\ with Old, Noninteractive Semantics}\label{sec:logic1}

The extended language is a version of Horn clauses
 with additive goals. It is described
by $G$- and $D$-formulas given by the syntax rules below:
\begin{exmple}
\>$G ::=$ \>  $A \sep   G \otimes  G \sep    \some x\ G \sep  G \adc G \sep  G \add G $ \\   \\
\>$D ::=$ \>  $A  \sep G \supset A\ \sep \all x\ D $\\
\end{exmple}
\noindent
In the rules above, $A$  represents an atomic formula.
A $D$-formula  is called a  Horn
 clause with additive goals, or simply a clause. 

In the transition system to be considered, $G$-formulas will function as 
queries and a set of $!D$-formulas will constitute  a program. 
 We will  present the standard operational 
semantics for this language  as inference rules \cite{Khan87}. 
The rules for executing queries in our language are based on
uniform provability \cite{HM94,MNPS91}. Below the notation $D;\Pscr$ denotes
$\{ D \} \cup \Pscr$ but with the $D$ formula being distinguished
(marked for backchaining). Note that execution  alternates between 
two phases: the goal reduction phase (one  without a distinguished clause)
and the backchaining phase (one with a distinguished clause).

\begin{defn}\label{def:semantics}
Let $G$ be a goal and let $\Pscr$ be a program that consists of only reusable clauses.
Then the task $\pr(\Pscr,G)$ is defined as follows:

\begin{numberedlist}

\item  $\pr(A;\Pscr,A)$. \% This is a success.

\item    $\pr((G_0\supset A);\Pscr,A)$ if 
 $\pr(\Pscr, G_0)$.

\item    $\pr(\all x D;\Pscr,A)$ if   $\pr([t/x]D;
\Pscr, A)$.

\item    $\pr(\Pscr,A)$ if   $D \in \Pscr$ and $\pr(D;\Pscr, A)$.

\item $\pr(\Pscr, G_0 \otimes G_1)$  if $\pr(\Pscr, G_0)$ and 
$\pr(\Pscr, G_1)$. 

\item $\pr(\Pscr, G_0 \adc G_1)$  if $\pr(\Pscr, G_0)$ and 
$\pr(\Pscr, G_1)$.

 \item $\pr(\Pscr, G_0 \add G_1)$  if $\pr(\Pscr, G_0)$.

 \item $\pr(\Pscr, G_0 \add G_1)$  if $\pr(\Pscr, G_1)$.

 \item $\pr(\Pscr, \some x G_0)$  if $\pr(\Pscr, [t/x]G_0)$.

\end{numberedlist}
\end{defn}
\noindent  
The above rules are based on the focused proof theory of linear logic.

\section{The \muprolog\ with New, Interactive Semantics}\label{sec:logic}

Again, the new rules of \muprolog\ are formalized by means of what it means to
execute a goal  $G$ from the program $\Pscr$.

\begin{defn}\label{def:semantics}
Let $G$ be a goal and let $\Pscr$ be a program as before.  
Then   executing $G$ from $\Pscr$ -- written as $\prove(\Pscr,G)$ --
 is defined as follows: 

\begin{numberedlist}

\item  $\prove(A;\Pscr,A)$. \% This is a success.

\item    $\prove((G_0\supset A);\Pscr,A)$ if 
 $\prove(\Pscr, G_0)$.

\item    $\prove(\all x D;\Pscr,A)$ if   $\prove([t/x]D;
\Pscr, A)$.

\item    $\prove(\Pscr,A)$ if  $D\in \Pscr$ and  $\prove(D;\Pscr, A)$.

\item $\prove(\Pscr,G_0 \otimes G_1)$ if $\prove(\Pscr,G_0)$  and 
  $\prove(\Pscr,G_1)$. Thus, the two goal tasks  must be done in parallel and both tasks 
must succeed for the current task to succeed.

\item $\prove(\Pscr,\exists x G_0)$  if  $\prove(\Pscr,[t/x]G_0)$. Typically, selecting the true term 
                      can be achieved via the unification process.

\item $\prove(\Pscr, G_0\adc G_1)$ if $read(k)$ and $\prove(\Pscr, G_i)$ and  $\pr(\Pscr,G_j)$ where  $i\ (=\ 0\ {\rm or}\ 1))$ is chosen by the user (stored in $k$) and $j$ is $(i+1)\ mod\ 2$.

\item $\prove(\Pscr, G_0\add G_1)$ if  $\prove(\Pscr, G_i)$  where  $i\ (=\ 0\ {\rm or}\ 1))$ is chosen by the machine.

\end{numberedlist}
\end{defn}

\noindent  
In the above rules, the symbols $\add, \adc$  provides 
choice operations: in particular, the symbol $\add$  allows for the mutually exclusive 
execution 
of goals \cite{Post}.

The operational notion of execution defined above 
is intuitive enough and the following theorem -- whose proof is rather obvious from the discussion in 
 \cite{HM94} and can be
shown using an induction on the length of derivations --
 shows the connection
between the old operational semantics of Lolli \cite{HM94} and the new operational semantics.

\begin{theorem}
Let $\Pscr$ be a program and $G$ be a goal in \muprolog. Executing $\lb\Pscr,G\rb$ terminates
with a success if and only if $G$ follows from $\Pscr$ in  intuitionistic  linear logic.
Furthermore, every  successful execution respects the declarative readings of the logical
connectives.
\end{theorem}

\section{Examples }\label{sec:modules}

As an  example, let us consider the following database which contains the today's flight
information for major airlines such as Panam and Delta airlines.

\begin{exmple}
\% panam(source, destination, dp\_time, ar\_time) \\
\% delta(source, destination, dp\_time, ar\_time) \\
$panam(paris, nice, 9:40, 10:50).$\\
$panam(nice, london, 9:45, 10:10).$\\
$delta(paris, nice, 8:40, 09:35).$\\
$delta(paris, london, 9:24, 09:50).$\\
\end{exmple}
\noindent Consider a goal $\some dt \some at\ panam(paris,nice,dt,at)$ $\adc$ $\some dt \some at\ 
delta(paris,nice,$ \\ $dt,at)$. This goal expresses the task of diagnosing whether the user has a choice between
Panam and Delta to fly from Paris to Nice today. Note that this goal is solvable because the user indeed 
does have a choice in the
example above.  The system in Section 2 requests the user to select a particular airline. After the airline -- 
 say, Panam -- is selected, the system produces the departure and arrival
time of the flight of the Panam airline, \ie, $dt = 9:40, at = 10:50$.

As another  example, let us consider the following database which contains tuition
information for some university. The following tuition and fee charges are in effect for 
this year: \$40K for full-time students, \$20K for part-time.

\begin{exmple}
$fulltime(40000)$.\\
$parttime(20000)$.\\
\end{exmple}
\noindent Consider a goal $\some x\ fulltime(x)  \adc \some x\ parttime(x)$. 
This goal expresses the task of diagnosing whether the user
is a full-time student or a part-time.   The system in Section 2  requests the user to select the current 
status.  After the status -- 
 say, full-time -- is selected, the system 
produces the amount, \ie, $x = 40K$.

\section{An Alternative Operational Semantics}\label{sec:0627}

Our execution model in the previous section  requires only small changes to the existing Prolog model and,
therefore, quite efficient. However, it has a serious drawback: it requests the user to perform some actions in advance
even when an execution leads to a failure. This feature is not appealing at all.
Fixing this problem requires fundamental changes to our execution model.

To be precise, the new execution model -- adapted from \cite{Jap03} -- requires two phases:

\begin{numberedlist}

\item the proof phase: This phase builds a proof tree. This proof tree 
 encodes how to solve a goal given a program.

\item the execution phase: This phase actually solves the goal relative to the program using the proof tree.

\end{numberedlist}

\noindent

Given a program $\Pscr$ and a goal $G$, a proof tree of $\Pscr \supset G$ is a list of tuples of
the form $\lb F,i \rb$ or $\lb F,(i,j) \rb$ where $F$ is a formula and $i,j$ are the distances to $F$'s chilren
in the proof tree. Below, $a_1::\ldots::a_n::nil$ represents a list of $n$ elements.

\begin{defn}\label{def:semantics}
Let $G$ be a goal and let $\Pscr$ be a program that consists of only reusable clauses.
Then the task of proving $\Pscr\supset G$ and returning its proof tree $L$ -- 
written as $\pr(\Pscr\supset G,L)$ -- is defined as follows:

\begin{numberedlist}

\item  $\pr(A;\Pscr\supset A,\lb A;\Pscr\supset A, - \rb::nil)$. \% This is a leaf node.

\item    $\pr((G_0\supset A);\Pscr\supset A, \lb (G_0\supset A);\Pscr\supset A,1\rb::L)$ if 
 $\pr(\Pscr\supset G_0,L)$.

\item    $\pr(\all x D;\Pscr\supset A, \lb \all x D;\Pscr\supset A,1\rb::L) $ if   $\pr([t/x]D;
\Pscr\supset A,L) $.

\item    $\pr(\Pscr\supset A,\lb \Pscr\supset A,1\rb::L)$ if   $D \in \Pscr$ and $\pr(D;\Pscr\supset A,L)$.

\item $\pr(\Pscr\supset G_0 \otimes G_1,\lb \Pscr\supset G_0 \otimes G_1,(m+1,1) \rb::L_2 )$ 
 if $\pr(\Pscr\supset G_0,L_0)$ and 
$\pr(\Pscr\supset G_1,L_1)$ and $append(L_0,L_1,L_2)$ and  $length(L_1,m)$. 

\item $\pr(\Pscr\supset G_0 \adc G_1,\lb \Pscr\supset G_0 \adc G_1,(m+1,1) \rb::L_2 )$ 
 if $\pr(\Pscr\supset G_0,L_0)$ and 
$\pr(\Pscr\supset G_1,L_1)$ and $append(L_0,L_1,L_2)$ and $length(L_1,m)$.

 \item $\pr(\Pscr\supset G_0 \add G_1,\lb \Pscr\supset G_0 \add G_1,1\rb::L)$  if $\pr(\Pscr\supset G_0,L)$.

 \item $\pr(\Pscr\supset G_0 \add G_1,\lb \Pscr\supset G_0 \add G_1,1\rb::L)$  if $\pr(\Pscr\supset G_1,L)$.
 
 \item $\pr(\Pscr\supset \some x G,\lb \Pscr\supset \some x G,1\rb::L)$  if $\pr(\Pscr\supset [t/x]G ,L)$.

\end{numberedlist}
\end{defn}
\noindent

Once a proof tree is built, the execution phase actually solves the goal relative to the program using the proof tree.
Below, the notation $A$ choose $B$ represents that the machine chooses one between two tasks $A$ and $B$.

\begin{defn}\label{def:exec}

Let $i$ be an index and let $L$ be a proof tree.  
Then   executing the $i$ element in $L$ -- written as $\prove(i,L)$ --
 is defined as follows: 

\begin{numberedlist}

\item  $\prove(i,L)$ if $member(i,L,T)$ and $T = (F,-)$. \% no child

\item  $\prove(i,L)$ if $member(i,L,T)$ and $T = (F,m)$  and $\prove(i-m,L)$. \% single child

\item  $\prove(i,L)$ if $member(i,L,T)$ and $T = (\Pscr\supset G_0 \otimes G_1,(n,m))$ and
 $\prove(i-n,L)$ and $\prove(i-m,L)$. \% two children

\item  $\prove(i,L)$ if $member(i,L,T)$ and $T = (\Pscr\supset G_0 \adc G_1,(n,m))$ and $read(k)$ and
 (($(k = 0)$ and $\prove(i-n,L)$) choose ($(k = 1)$ and $\prove(i-m,L)))$. 

 Here, the machine requests the user to type 0 or 1 and executes the chosen path.

\end{numberedlist}
\end{defn}

\noindent  
Now given a program $\Pscr$ and a goal $G$, solving $G$ from $\Pscr$ -- $exec(\Pscr,G)$ -- is defined as the following:

\[ \pr(\Pscr\supset G,L)\ {\rm and}\ length(L,n)\ {\rm and}\ \prove(n,L).  \]

\section{Conclusion}\label{sec:conc}

In this paper, we have considered an extension to Prolog with  
additive  goals in linear logic. This extension allows goals of 
the form  $G_0 \add  G_1$  and $G_0 \adc  G_1$ where $G_0, G_1$ are goals.
In particular, the latter goals make it possible for  Prolog
to model decision steps from the user.

We plan to connect our execution model to Japaridze's Computability Logic \cite{Jap03,Jap08}
 in the near future.



\bibliographystyle{plain}



\end{document}